\documentclass{aa}
\usepackage[applemac]{inputenc}
\usepackage{natbib}
\usepackage{txfonts}
\usepackage{graphicx}
\bibpunct{(}{)}{;}{a}{}{,}
\def\Arrow{\mathop{\longrightarrow}\limits}

\begin{document}

\title{Modelling line emission of deuterated H$_3^+$ from prestellar cores}
\author{O. Sipilä\inst{1}
\and{E. Hugo\inst{2}}
\and{J. Harju\inst{1}}
\and{O. Asvany\inst{2}}
\and{M. Juvela\inst{1}}
\and{S. Schlemmer}\inst{2}}
\institute{Observatory, PO Box 14, 00014, University of Helsinki, Finland\\
e-mail: \texttt{olli.sipila@helsinki.fi}
\and{I. Physikalisches Institut, Universität zu Köln, Germany}}

\abstract{The depletion of heavy elements in cold cores of interstellar molecular clouds can lead to a situation where deuterated forms of ${\rm H_3^+}$ are the most useful spectroscopic probes of the physical conditions.}{The aim is to predict the observability of the rotational lines of ${\rm H_2D^+}$ and ${\rm D_2H^+}$ from prestellar cores.}{Recently derived rate coefficients for the ${\rm H_3^+} +{\rm H_2}$ isotopic system were applied to the "complete depletion" reaction scheme to calculate abundance profiles in hydrostatic core models. The ground-state lines of ${\rm H_2D^+(o)}$ (372 GHz) and ${\rm D_2H^+(p)}$ (692 GHz) arising from these cores were simulated. The excitation of the rotational levels of these molecules was approximated by using the state-to-state coefficients for collisions with ${\rm H_2}$. We also predicted line profiles from cores with a power-law density distribution advocated in some previous studies.}{The new rate coefficients introduce some changes to the complete depletion model, but do not alter the general tendencies. One of the modifications with respect to the previous results is the increase of the ${\rm D_3^+}$ abundance at the cost of other isotopologues. Furthermore, the present model predicts a lower H$_2$D$^+$ (o/p) ratio, and a slightly higher D$_2$H$^+$ (p/o) ratio in very cold, dense cores, as compared with previous modelling results. These nuclear spin ratios affect the detectability of the submm lines of H$_2$D$^+$(o) and D$_2$H$^+$(p). The previously detected H$_2$D$^+$ and D$_2$H$^+$ lines towards the core I16293E, and the H$_2$D$^+$ line observed towards Oph D can be reproduced using the present excitation model and the physical models suggested in the original papers.}{}


\maketitle

\section{Introduction}

The rate coefficients for reactive and inelastic collisions between the H$_3^+$ ion and H$_2$ with all possible deuterated variants and nuclear spin symmetries have been recently derived and a few of them have been experimentally tested by \citet{HAS09}. Besides making state-specific astrochemical models viable, the cross-sections derived in this work make it possible to model the intensities of the dipole-allowed rotational transitions of H$_2$D$^+$ and D$_2$H$^+$.

The H$_2$D$^+$ and D$_2$H$^+$ ions are potentially useful probes of pre-protostellar cores \citep[see e.g.][]{Caselli03, Vastel04, VanderTak05, Pagani09}. Together with the isotopologues H$_3^+$ and D$_3^+$, they are likely to belong to the most abundant ions in the dense and cold nuclei of prestellar cores, where the usual tracer molecules can have practically disappeared from the gas phase \citep{Roberts03, WFP04} -- this state of matters is called ``complete depletion'' in the latter paper. While it has been observed that heavier substances such as CN can exist in the gas phase at densities of the order of $10^6$ cm$^{-3}$ \citep{HWPF08}, the assumption of complete depletion remains a valid first approximation at high densities.

The astrophysical significance of these ions is related to the facts that H$_3^+$ originates almost directly in the cosmic ray ionization of H$_2$, and that it initiates the ion-molecule reactions in dense clouds \citep{Herbst73}. Furthermore, the deuteration of H$_3^+$ depends on the ortho-para ratio of H$_2$, which is likely to be non-thermal and evolve with time in interstellar clouds \citep{Pagani92, Gerlich02, FPW06, Pagani09}.

In this paper we apply the newly derived chemical rate coefficients to the "completely depleted" case first discussed by \citet{WFP04}, and use the state-to-state coefficients to predict the ground-state rotational lines of {\sl ortho}-H$_2$D$^+$ and {\sl para}-D$_2$H$^+$ from prestellar cores. The two core models used in these simulations correspond to the observed properties of the cores Oph D and I16293E in Ophiuchus. The rate coefficients of \citet{HAS09} for some important deuteration reactions differ from those adopted in \citet{Roberts03}, and in the series of papers by Flower, Pineau des Forêts \& Walmsley, and we discuss changes implied to the complete depletion model. The \citet{HAS09} rate coefficients have been previously used by \citet{Pagani09} in the modelling of the prestellar core L183, where also CO, N$_2$ and their derivatives were included in the chemical reaction scheme. The organization of the present paper is as follows: in Sect. \ref{sect2}, we describe the physical and chemical models and, in Sect. \ref{sect3}, we present the modelling results. The results are discussed in Sect. \ref{sect4}, and finally, the main points of this discussion are summed up in Sect. \ref{sect5}.

\section{Model}\label{sect2}

In what follows, we describe the model assumptions. These include the physical core model, the description of the dust grain component, the chemical reaction scheme, and the adopted rate coefficients. We also give a brief account of the methods and programs used in solving the chemical abundances and the emitted molecular line radiation.

\subsection{Core model}

For the core model we used a modified Bonnor-Ebert sphere (BES), which is a non-isothermal cloud in hydrostatic equilibrium \citep[previously discussed by, e.g.,][]{Evans01}, heated externally by the interstellar radiation field, ISRF. The possibility of a hydrostatic or near-equilibrium configuration is suggested by observations towards some prestellar cores which seem to represent an advanced stage of chemical evolution (characterized by a high degree of molecular depletion), and have near-thermal linewidths \citep[e.g.,][]{Bergin07}.

We first calculated a density profile resulting from an isothermal model \citep{Bonnor56}. The density profile was then fed into a Monte Carlo radiative transfer program \citep{JP03, Juv05} for the temperature calculation. For this purpose the core model was divided into concentric shells.  We used the grain size distribution of \citet{MRN77} (MRN) with two types of dust grains: carbonaceous and silicate (the grain model is further discussed in Sect. \ref{grains}). The choice of using the MRN distribution implies that the abundance of very small grains (with radii less than $\sim$ 0.01 $\mu$m) is assumed to be negligible, and that grain coagulation is not considered.  The temperature profile was calculated separately for the two dust types. In each position of the core, the final temperature profile was taken to be the arithmetic average of the temperatures of both grain types.

The gas and dust temperatures were assumed to be equal. While this is probably a good approximation in the dense ($> 10^6\,{\rm cm^{-3}}$) central part of the core \citep{Burke83}, it may not hold as well in the outer layers of the core \citep[$\sim 10^5\,{\rm cm^{-3}}$,][]{Bergin06}. In the present analysis, which concentrates on the very dense nucleus of the core, we have ignored this phenomenon. The model core is assumed to be embedded in a molecular cloud with extinction corresponding to $A_{\mathrm V} = 10^{\rm m}$. This assumption agrees with observations from the vicinity of Oph D \citep{Chapman09}.

To determine the exact core density profile using the computed temperature distribution, we slightly modified the method used in \citet{Bonnor56} by making the following substitutions:
\begin{equation}
\rho = \lambda\left({T_{\mathrm c}\over{T}}\right){\mathrm e}^{-\psi}
\end{equation}
\begin{equation}
r = \beta^{1/2}\lambda^{-1/2}\xi,
\end{equation}
where $\beta = k_{\mathrm B}T_{\mathrm c}/{4\pi{Gm}}$, $T_{\mathrm c}$ is the temperature in the center of the core and $m$ is the mass of the hydrogen molecule. As in \citet{Bonnor56}, $\lambda$ was chosen to be the central density. We also imposed the boundary conditions $\psi = 0$, ${\mathrm d}\psi/{\mathrm d}\xi = 0$ and ${\mathrm d}T/{\mathrm d}\xi = 0$ in the centre of the core. Substituting the above expressions into the equation of hydrostatic equilibrium (Bonnor's equation 2.2) and keeping in mind that $T = T(r)$, one obtains
\begin{equation}\label{1}
{{\mathrm d}^2\psi\over{\mathrm d}\xi^2} = \left({T_{\mathrm
c}\over{T}}\right)^2{\mathrm e}^{-\psi} - {2\over\xi}{{\mathrm
d}\psi\over{\mathrm d}\xi} - {1\over{T}}{{\mathrm d}T\over{\mathrm
d}\xi}{{\mathrm d}\psi\over{\mathrm d}\xi} \, .
\end{equation}
Equation (\ref{1}) was integrated numerically and the resulting density profile was fed back into the radiative transfer program to calculate a new temperature profile. This procedure was repeated a few times until the density profile converged.

\subsection{Chemistry model}

We adopted the complete depletion model discussed in \citet{WFP04} and \citet{FPW04}. The reaction scheme is fairly simple: the gas-phase reactants include only H, H$^+$, H$_2$, H$_2^+$, H$_3^+$ and their deuterated forms, He, He$^+$, and free electrons. Because of the low temperature associated with the environment, the zero-point energies become relevant and thus the deuterated forms and different nuclear spin modifications have to be considered explicitly. The reaction set also includes cosmic ray ionization, H$_2$ formation on grains, and some other grain processes. We next discuss the gas-phase and grain-surface reactions separately.

\subsubsection{Gas phase chemical reactions}

In all our models, we assumed that the core is electrically neutral, i.e. that electron abundance equals the difference between the total abundance of positive ions and the abundance of negatively charged grains.

The chemical reactions and the associated rate coefficients were compiled using four sources: \citet{WFP04}, \citet{FPW04}, \citet{HAS09}, and \citet{Pagani09}. The reaction set of \citet{WFP04} was complemented by \citet{FPW04} to include the nuclear spin modifications of ${\rm D_2}$, ${\rm D_2H^+}$, and ${\rm D_3^+}$. \citet{HAS09} took the analysis of the H$_3^+$ + H$_2$ reacting system further by using a microcanonical approach to derive state-to-state rate coefficients. This approach differs from that of \citet{FPW04} in that all internal states of the reacting system are explicitly taken into account, resulting in detailed state-to-state rate coefficients for different configurations of the system. We adopted their ground state-to-species rate coefficients for all reactions relevant to the H$_3^+$ + H$_2$ system (including the deuterated forms) and substituted them into the combined data of \citet{WFP04} and \citet{FPW04}. Furthermore, for dissociative recombination (DR) reactions of ${\rm H_3^+}$ and its deuterated forms we used the newly calculated rate coefficients presented in Appendix B of \citet{Pagani09}.

\subsubsection{Grain reactions}

The H$_2$ and HD molecules are mainly formed on grain surfaces \citep{GS63, HS71}, whereas for the formation of D$_2$, gas-phase reactions are more important. Nevertheless, the formation of all three isotopologues on dust grains, and the corresponding destruction terms for atomic H and D were included in the chemical system. We assumed that {\sl ortho} and {\sl para} forms are produced according to their statistical population ratios 3:1 and 2:1 for H$_2$ and D$_2$, respectively. 

Grain processes, in particular the formation of {\sl ortho}-H$_2$ and the attachment and destruction of positive ions on grain surfaces, are important for the overall degree of deuteration of the cloud (see Sect. \ref{sect4}). The details of these processes (grain surface properties, desorption mechanisms, etc.) were not considered in the present model. It was assumed that chemical reactions on grains (along with the associated adsorption and desorption) take place instantaneously.

The rate coefficients for grain reactions (including, e.g., electron attachment and recombination of positive ions on grains), listed in \citet{WFP04}, were taken from \citet{FP03}, in which an MRN distribution with $a_{\mathrm {min}} = 0.01$ $\mu$m, $a_{\mathrm {max}} = 0.3$ $\mu$m was assumed. These limits correspond to an effective grain radius of $a_{\rm eff} = 0.0202$ $\mu$m\footnote{The rate coefficients of grain reactions listed in Table A.1. of \citet{WFP04} correspond to a$_{\rm eff}$ = 0.02 $\mu$m, contrary to what is said in the caption.}.  Since the rate coefficient for a grain reaction is proportional to $\sigma_{\mathrm g} = \pi{a_{\rm eff}^2}$, we scaled the \citet{FP03} coefficients by $(a_{\rm eff}/0.02 \mu{\rm m})^2$, where $a_{\rm eff}$ is the effective grain radius of the assumed size distribution. The Coulomb factor, $\tilde{J}(\tau, \nu)$, which takes into account the electric interaction between dust grains and gas phase neutrals and/or ions, was included as an additional factor. Its mathematical form has been discussed in detail in \citet{DS87}, and recently in \citet{Pagani09}.

\subsubsection{Molecular line radiation}

The abundance distributions of chemical species were determined by a chemistry program in physical conditions corresponding to those in different parts of the model core -- in practice in each of the concentric shells used in the temperature calculations.  The radial ${\rm H_2D^+}$(o) and ${\rm D_2H^+}$(p) abundance distributions, together with the density and temperature profiles were used as input for a Monte Carlo radiative transfer program \citep{Juv97} to predict observable line emission.

The excitation of the rotational transitions of ${\rm H_2D^+}$(o) and ${\rm D_2H^+}$(p) in collisions with {\sl para} and {\sl ortho} H$_2$ were calculated using the state-to-state rate coefficients from \citet{HAS09}.  The data concerning the line frequencies and Einstein $A$ coefficients were obtained from \citet{Miller89}, \citet{Ramanlal04}, and \citet{Amano05}.

\begin{figure}
\resizebox{\hsize}{!}{\includegraphics{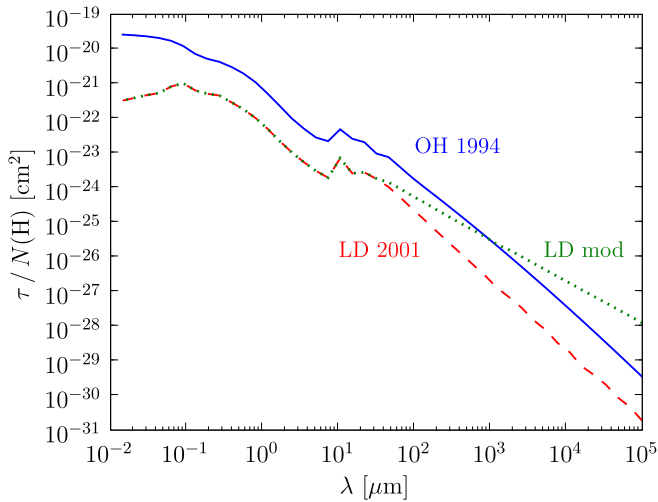}}
\caption{Optical thickness per column density according to the opacity data of \citet{OH94}, denoted in the figure by 'OH 1994', (solid curve) and \citet{LD01}, denoted by 'LD 2001'  (dashed curve). The modified extinction curve (see text) is also plotted, denoted by 'LD mod' (dotted curve).}
\label{fig: thickness}
\end{figure}

\subsubsection{Dust grain model}\label{grains}

The MRN model was adopted for the grain size distribution. In the density and temperature profile calculations, we used two types of grains; carbonaceous and silicate. The optical properties of these
components were adopted from \citet{LD01}. The densities of the carbonaceous and silicate grains were taken to be $2.5$ g\,cm$^{-3}$ and $3.5$ g\,cm$^{-3}$, respectively. In the chemistry model, no separation between the grain materials was done, and a grain material density of $3.0$ g\,cm$^{-3}$ was assumed. For the comparison with the \citet{FPW04} results (Sect. 3.1) we used, however, the same assumptions of the grain material density ($2.0$ g\,cm$^{-3}$) and dust-to-gas mass ratio (0.013) as in \citet{WFP04}.

\begin{figure}
\resizebox{\hsize}{!}{\includegraphics{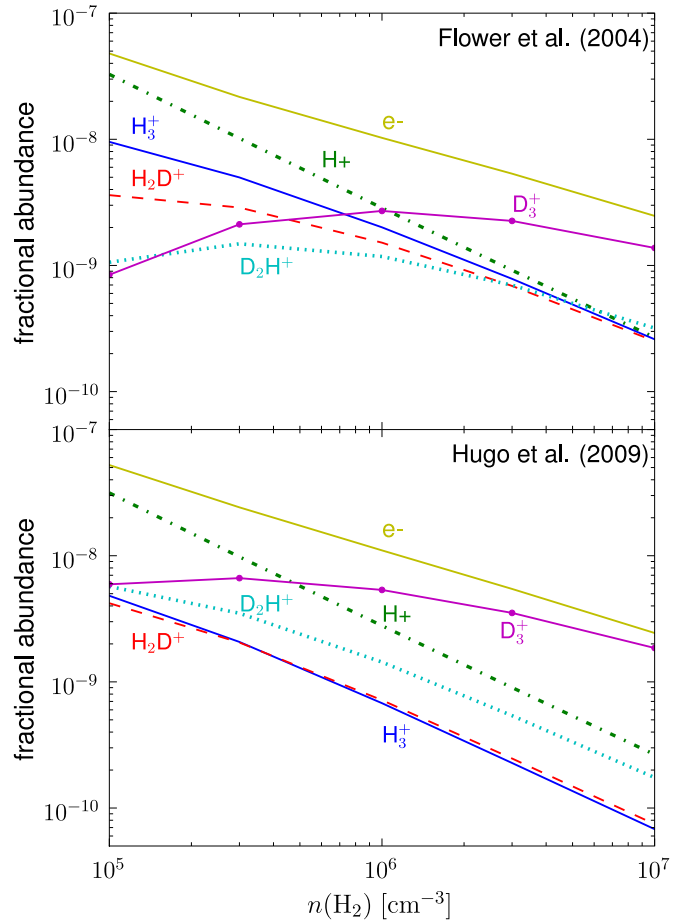}}
\caption{Steady state abundances of electrons, protons, H$_3^+$ and its deuterated forms, calculated using the \citet{FPW04} (upper panel) and \citet{HAS09} (lower panel) rate coefficients in an isothermal ($T = 10$\,K) model with single size grains ($a = 0.1$\,$\mu$m). A cosmic ray ionization rate of $\zeta=3\times10^{-17}$ s$^{-1}$ is assumed.}
\label{fig: fpwhas comparison}
\end{figure}

The model of \citet{LD01} describes dust in diffuse medium. In dense clouds, coagulation and growth of icy grain mantles are expected to modify the grain properties and, according to the models of \citet{OH94}, the extinction curve is expected to become flatter in the far-infrared and at longer wavelengths. Since in our cores the central densities are high, $n({\rm H}) \sim 10^6$ cm$^{-3}$, we modified the 
\citet{LD01} dust opacities accordingly. This leads to temperatures of the order of 6 K in the core center. The curves are plotted in Fig. \ref{fig: thickness}.

\section{Results}\label{sect3}

In this Section we discuss the differences arising from the use of the modified chemical reaction rate coefficients used in this paper compared to those used in \citet{FPW04}. We then present results of chemical modelling carried out using a hydrostatic core model. We conclude the section with a brief discussion on simulated line emission spectra.

\subsection{Comparison with \citet{FPW04} results using homogeneous models}

\begin{figure}
\resizebox{\hsize}{!}{\includegraphics{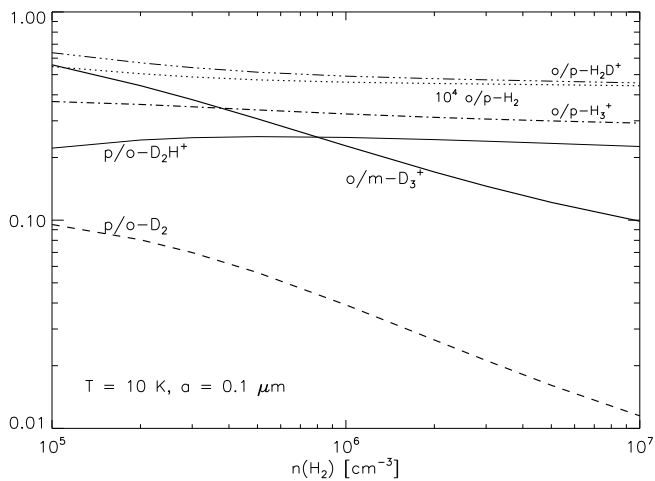}}
\caption{Steady-state abundance ratios of the nuclear spin modifications of H$_2$, D$_2$, H$_3^+$ and its deuterated forms as functions of $n({\rm H_2})$ calculated using \citet{HAS09} rate coefficients.  The model parameters are the same as used in Fig.~\ref{fig: fpwhas comparison}.}
\label{fig: op_vs_dens}
\end{figure}

\begin{figure}
\resizebox{\hsize}{!}{\includegraphics{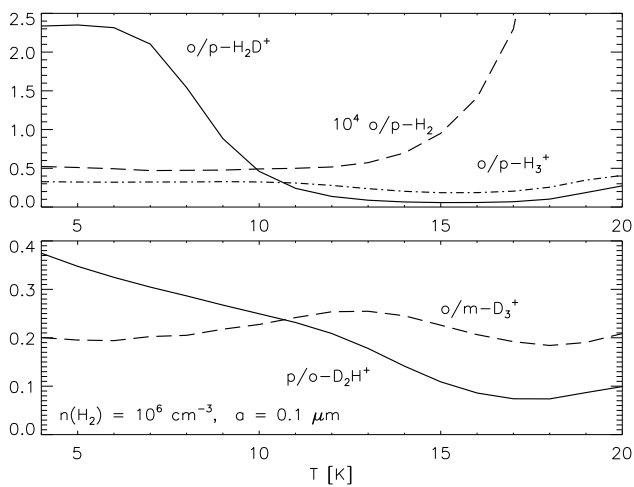}}
\caption{Steady-state o/p ratios of ${\rm H_2}$ (multiplied by $10^4$), ${\rm H_3^+}$, and ${\rm H_2D^+}$ (top panel), and the steady-state ${\rm D_2H^+}$ (p/o) and ${\rm D_3^+}$ (m/o) ratios (bottom panel) as functions of the gas temperature in the range $4-20$ K. The assumed gas density and grain radius are $n({\rm H_2}) = 10^6$ cm$^{-3}$ and $a=0.1 \, \mu$m, respectively.}
\label{fig: op_vs_t}
\end{figure}

We ran a set of homogeneous models corresponding to those presented in \citet{WFP04} and in \citet{FPW04}. The steady state fractional abundances of the principal ions as functions of the gas density are shown in Fig. \ref{fig: fpwhas comparison}. The upper diagram corresponds to the \citet{FPW04} model and is a reproduction of their Fig. 1.  The lower plot is produced using the \citet{HAS09} rate coefficients for the H$_3^+$ + H$_2$ isotopic system and the DR rate coefficients from \cite{Pagani09}. The figures were produced assuming an isothermal ($T = 10$\,K) model with single sized grains ($a = 0.1$\,$\mu$m) and a grain material density of 2.0 g\,cm$^{-3}$.

The abundance ratios of the most important nuclear spin variants as functions of the gas density are shown in Fig.~\ref{fig: op_vs_dens}, and the temperature dependence at low temperatures is illustrated in Fig.~\ref{fig: op_vs_t}.  The former should be compared with Fig. 2 of \cite{FPW04} and with Fig. 3 of \cite{WFP04}, and the latter with Figs. 5 and 6 in \citet{FPW04}. Following \cite{HAS09} we have assigned {\sl meta}-D$_3^+$ with the modification having the lowest ground state energy, corresponding to the $A_1$ representation of the symmetry group $S_3$, and {\sl ortho}-D$_3^+$ with the $E$ representation (see their Sect. II.A.1). The {\sl ortho} and {\sl meta} appelations for ${\rm D}_3^+$ are therefore interchanged with respect to those used in \cite{FPW04}, \cite{Pagani09}, and most other papers. As in \citet{FPW04}, the D$_3^+$ (o/m) and D$_2$ (p/o) ratios fall rapidly with increasing $n({\rm H_2})$.  However, the ${\rm D_2H^+}$ (p/o) ratio is nearly constant unlike in the previous study. The H$_3^+$ (o/p) ratio is about three times lower in the present model. There are also a couple of differences in the temperature dependence (Fig. 4). The H$_2$D$^+$ (o/p) ratio levels off between 2-3 when approaching very low temperatures. Using the \citet{FPW04} rate coefficients the corresponding value is $\sim$ 10 in the same conditions. Secondly, the D$_2$H$^+$ (p/o) reaches slightly higher $(0.3 - 0.4)$ at very low temperatures than in the \citet{FPW04} model. We note that as our model follows the chemistry until steady state, the H$_2$ (o/p) ratio becomes much lower than in the models of \citet{Pagani09}. As a consequence, we obtain a clearly higher degree of deuteration of H$_3^+$ and somewhat different nuclear spin ratios for its isotopologues as compared with this previous study. We return to this matter briefly in Sect. 4.1.

All the plots presented so far assume single sized grains with a radius of 0.1\,$\mu$m. As discussed in Sect. 3.3 of \cite{FPW04}, the grain size has a considerable effect on the abundances and nuclear spin ratios. The D$_2$ (o/p), D$_2$H$^+$ (o/p), and D$_3^+$ (m/o) ratios peak near the grain radius $a= 0.1\, \mu$m \citep[see Fig. 9 in][]{FPW04}, whereas the H$_2$ (o/p), H$_3^+$ (o/p), and H$_2$D$^+$ (o/p) ratios increase monotonically towards smaller grain radii.

\subsection{Hydrostatic models}

We studied the influence of the grain size distribution and the cosmic ionization rate on the chemistry of a depleted, hydrostatic core. The model BE sphere was assumed to have a central density of $n(\mathrm{H}_2)$ $= 2\times10^6$ cm$^{-3}$ and an outer radius of 2400 AU, where the density drops to about $10^5$ cm$^{-3}$. These parameters correspond roughly to the dense nucleus of the Oph D core according to \cite{Harju08}. The effective grain radius was varied by changing the lower limit of the grain size distribution and keeping the upper limit constant at 0.3 $\mu$m. Modelling was carried out with effective grain radii of 0.05 $\mu$m, 0.1 $\mu$m and 0.2 $\mu$m. The slope of the extinction curve (Fig. \ref{fig: thickness}) was modified to produce a central core temperature of about 6 K (see Sect. \ref{grains}). The chemical reaction network was integrated until $t = 2\times10^6$ years, which was well within steady state (all the abundances settled into steady state at times $< 10^6$ years).

\begin{figure}
\resizebox{\hsize}{!}{\includegraphics{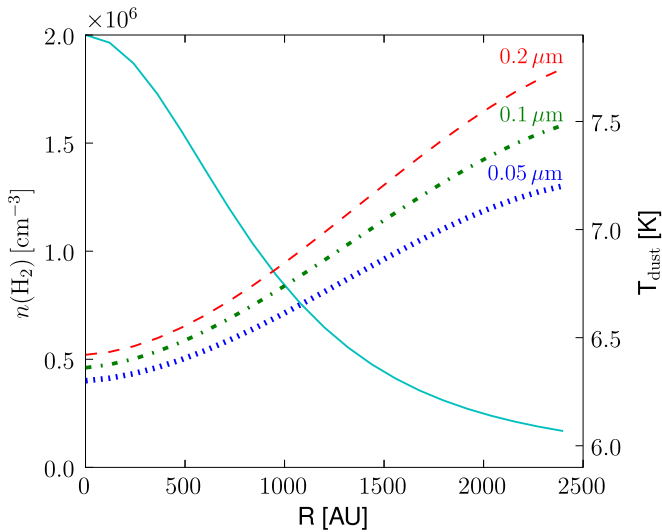}}
\caption{The radial core density profile corresponding to effective grain radius $a_{\mathrm {eff}}$ = 0.1 $\mu$m, with a central temperature of $\sim$ 6.4 K. Radial temperature profiles corresponding to different values of effective grain radius (indicated in the Figure) are superposed.}
\label{fig: density}
\end{figure}

Figure \ref{fig: density} shows the core density profile corresponding to effective grain radius $a_{\mathrm {eff}}$ = 0.1 $\mu$m. The temperature profiles for three different grain sizes are superposed. Changing $a_{\mathrm{eff}}$ has a marked effect on the temperature profile, but alters the density distribution only slightly (the central density is kept fixed in our models). The core gets colder as the effective grain radius gets smaller; this is because the total grain surface area (effectively the total opacity) grows larger as the effective grain radius decreases.

\begin{figure}
\resizebox{\hsize}{!}{\includegraphics{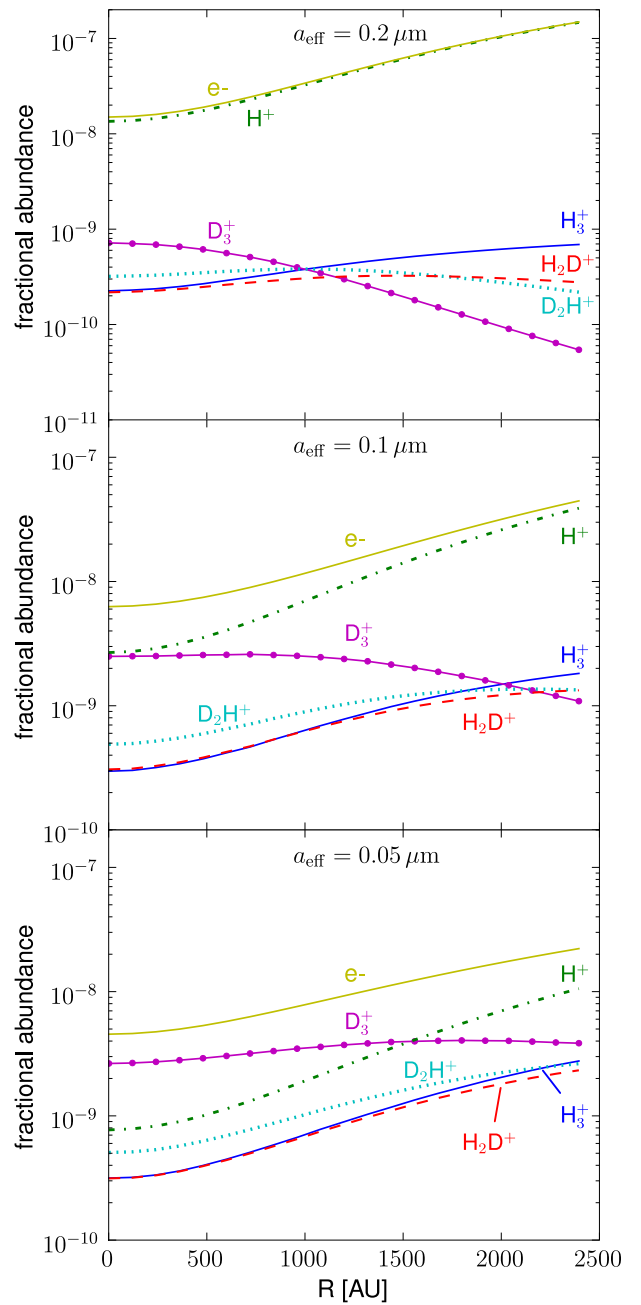}}
\caption{Radial fractional (with respect to $n({\mathrm {H_2}})$) adundance profiles of electrons, H$^+$, H$_3^+$ and its deuterated forms at $t = 1\times10^6$ years, for a cosmic ray ionization rate $\zeta = 3\times10^{-17}$\,s$^{-1}$.}
\label{fig: abundances_1.0}
\end{figure}

Figure \ref{fig: abundances_1.0} shows the steady state fractional abundance profiles of the principal ions for the selected values of effective grain radius and for a cosmic ray ionization rate of $\zeta = 3$ $\times$ $10^{-17}$\,s$^{-1}$. The grain radius decreases from top to bottom.  As shown in Fig. \ref{fig: density}, this decrease is accompanied with a slight drop in the average temperature. The fractional ion abundances generally decrease towards the core centre, i.e., towards higher densities, in accordance with the results from homogeneous models shown in Fig. \ref{fig: fpwhas comparison}. The density dependence of D$_3^+$ is less marked, which leads to an increased degree of deuterium fractionation in the centre when the effective grain radius decreases. The fractional H$^+$ abundance decreases drastically with decreasing grain size. This is caused by accentuated recombinations on negatively charged grains as explained in \cite{WFP04}.

\begin{figure}
\resizebox{\hsize}{!}{\includegraphics{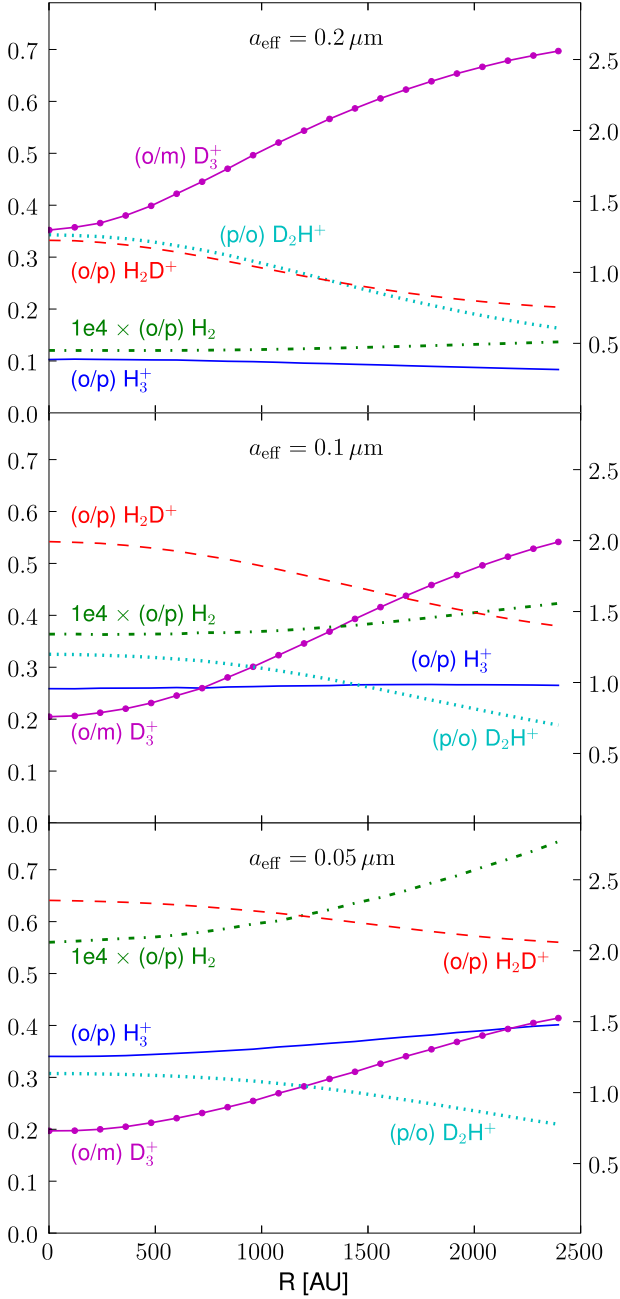}}
\caption{Radial profiles for the ratios of the different spin states of H$_2$, H$_3^+$ and its deuterated forms at $t = 1\times10^6$ years, for a cosmic ray ionization rate $\zeta = 3\times10^{-17}$\,s$^{-1}$. The y-axis on the right side denotes the H$_2$D$^+$ (o/p) ratio.}
\label{fig: opratios}
\end{figure}

The radial profiles of the ${\rm H_2}$ (o/p), ${\rm H_3^+}$ (o/p), ${\rm H_2D^+}$ (o/p), ${\rm D_2H^+}$ (p/o), and ${\rm D_3^+}$ (o/m) ratios are presented Figure \ref{fig: opratios}.  While most ratios have rather flat distributions, the ${\rm D_3^+}$ (o/m) drops towards the centre, i.e. towards higher densities. The density dependence is, however, smoother for small grain sizes (bottom panel). The ortho/para ratios of H$_2$, H$_3^+$, and H$_2$D$^+$ increase from top to bottom, along with the decreasing grain size.  This is caused by an intensified replenishment of H$_2$(o) when the total grain surface area becomes larger. The slight increase of H$_2$ (o/p) towards the outer parts reflects the decreasing density (see Fig. \ref{fig: op_vs_dens}). The ${\rm H_3^+}$ (o/p) ratio follows the H$_2$ (o/p) ratio closely. The ${\rm H_2D^+}$ (o/p) ratio correlates with the H$_2$ (o/p) ratio, but rises towards low temperatures (see Fig. \ref{fig: op_vs_t}). These tendencies result in a slight increase of the ${\rm H_2D^+}$ (o/p) ratio towards the core centre. The D$_2$H$^+$ (p/o) ratio changes very little as a function of radius and from model to model.  Like the D$_3^+$ (o/m) ratio, the D$_2$H$^+$ (o/p) ratio has a flat maximum around $a \sim 0.1\,  \mu$m \citep{FPW04}, and depends weakly on the density (Fig. \ref{fig: op_vs_dens}).

The effect of changes in the cosmic ionization rate, $\zeta$, is illustrated in Fig. \ref{fig: crp_comparison}.  This shows the radial distributions of the principal ions for the cases where $\zeta$ is made 10 times lower (top) and 10 times higher (bottom) than in the models discussed above. Lowering $\zeta$ decreases the degree of ionization. An enhanced cosmic ray ionization increases the ion abundances, but decreases the deuteration through an intensified electron recombination \citep{WFP04, FPW04}. The effect is particularly marked for ${\rm D_3^+}$ near the outer edge of the core.

\subsection{H$_2$D$^+$ and D$_2$H$^+$ spectra}\label{sect3spectra}

We calculated the spectral line profiles of the ground state transitions of ${\rm H_2D^+(o)}$ ($1_{10}-1_{11}$, 372 GHz) and ${\rm D_2H^+(p)}$ ($1_{10}-1_{01}$, 692 GHz) from models corresponding to the observed properties of the prestellar cores I16293E and Oph D.

The very narrow 372 GHz line of ${\rm H_2D^+(o)}$ detected with APEX towards the density maximum of Oph D has been suggested to originate in a hydrostatic core (Harju et al. 2008). In accordance with this suggestion the physical model used for Oph~D is a hydrostatic core heated externally by the interstellar radiation field (ISRF). Judging from previous maps, in particular the ISOCAM $7\,\mu$m image from \citet{Bacmann00}, we assumed an angular radius of $20\arcsec$. We adopted the recent distance estimate of 120 pc to the Ophiuchus complex \citep{Lombardi08}, which implies that the adopted radius corresponds to 2400 AU.  The obscuration provided by the surrounding molecular cloud was assumed to correspond to $A_{\rm V} = 10^{\rm m}$. We varied the central H$_2$ density ($n_0$), the effective grain radius ($a_{\rm eff}$), and the cosmic ray ionization rate ($\zeta$), and calculated the appropriate abundance profiles using the chemistry model described above. Finally, the populations of the rotational levels of ${\rm H_2D^+(o)}$, and the $1_{10}-1_{11}$ line profiles were calculated using the state-to-state rate coefficients for ${\rm H_2D^+(o) + H_2}$ (para and ortho separately) collisions from \citet{HAS09} and our Monte Carlo radiative transfer program.  Calculations with different values for the model parameters were carried out until a reasonable agreement with the observed ${\rm H_2D^+(o)}$ profile was met. This model was then used to predict the ${\rm D_2H^+(p)}$ line at 692 GHz as observed with APEX.

The results of our simulations are presented in Fig. \ref{fig: spectra}, which also shows the observed ${\rm H_2D^+(o)}$ spectrum. The best fit was obtained using the following parameters: $n_0 = 2 \times 10^6 \, {\rm cm^{-3}}$, $a_{\rm eff}=0.1\, \mu$m, and $\zeta=6\times10^{-17}\, {\rm s}^{-1}$. The model core has a central temperature $T = 6.36$\,K. The fractional H$_2$D$^+$(o) abundance is $\sim$ $3.6\times10^{-10}$ in the core center, rising to $\sim$ $7.2\times10^{-10}$ at the edge. A Gaussian fit to the model spectrum yields an FWHM of 0.35\,km\,s$^{-1}$; the peak optical depth is $\tau_{\mathrm {peak}}$ $\sim$ 1.5. A molecular line profile for the D$_2$H$^+$(p) $1_{10}$ -- $1_{01}$ transition using the same core model parameters is shown in the lower panel of the figure. A Gaussian fit to the spectrum yields an FWHM of 0.27\,km\,s$^{-1}$, the peak optical depth is $\tau_{\mathrm {peak}}$ $\sim$ 0.6. The D$_2$H$^+$(p) abundance is $\sim$ $1.4\times10^{-10}$ at the core center and $\sim$ $1.1\times10^{-10}$ at the edge -- essentially a flat profile with a minor maximum at around 1200 AU.

Both ${\rm H_2D^+}$ and ${\rm D_2H^+}$ have been detected with the CSO towards I16293E by \citet{Vastel04}. To our knowledge this work has produced the only rather firm detection of the 692 GHz line so far. \citet{Vastel04} estimated the ${\rm H_2D^+(o)}$ and ${\rm D_2H^+(p)}$ column densities from the observed line profiles by assuming line-of-sight homogeneity and an excitation temperature of $T_{\rm ex} = 10$ K.  We assumed that these column densities are approximately valid, and adjusted the fractional abundances accordingly. The physical model of the core was adopted from \citet{Stark04}, allowing for the fact that the cloud is probably nearer than previously thought \citep[120 pc,][]{Lombardi08}. The model consisted of a compact, homogeneous nucleus with a density $n_0 = 1.6 \, \times \, 10^6 \, {\rm cm^{-3}}$ up to a radius of 750 AU, surrounded by an envelope with the density power law $n \propto r^{-1}$ extending to a distance of 6000 AU ($50\arcsec$). A dust temperature $T_{\rm dust} = 16$ K derived by \citet{Stark04} was adopted as the temperature of the nucleus. The dust temperature was gradually increased toward the core edge, reaching a value $T_{\rm dust} = 20$ K at 6000 AU.  It should be noted, however, that the observations of \citet{Vastel04} suggest a lower value for the region where the ${\rm H_2D^+}$ and ${\rm D_2H^+}$ emission originate. Assuming thermal broadening only, the line widths listed in their Table 1 imply kinetic temperatures $T_{\rm kin} = 11.2\,\pm\,2.5$ (${\rm H_2D^+}$) and $T_{\rm kin} = 9.1\,\pm\,4.4$ (${\rm D_2H^+}$). These estimates are upper limits as they neglect the possible turbulent broadening and instrumental effects.  In order to reproduce the column densities derived by \citet{Vastel04}, the fractional ${\rm H_2D^+}$(o) abundance was set to $\sim 10^{-10}$ in the nucleus, and it was let to fall radially to $\sim 10^{-11}$ at the outer edge. The $\rm D_2H^+$(p) abundance was set at 0.6 times the ${\rm H_2D^+}$(o) abundance, which is within the column density error range reported by \citet{Vastel04}.

The line profiles calculated from this model are shown in Fig. \ref{fig: 16293_spectra}. The simulated H$_2$D$^+$(o) line has a peak optical depth $\tau$ $\sim$ 0.33. A gaussian fit to the spectrum yields an FWHM of $\sim$ 0.46 km\,s$^{-1}$ and a peak antenna temperature $T^*_{\rm A}$ $\sim$ 1.42 K. The simulated D$_2$H$^+$(p) line yields $\tau$ $\sim$ 0.30, an FWHM of $\sim$ 0.41 km\,s$^{-1}$ and a peak $T^*_{\rm A}$ $\sim$ 0.44 K. The antenna temperatures of the simulated H$_2$D$^+$(o) and D$_2$H$^+$(p) spectra agree well with the observations of \citet{Vastel04}, but the lines are too broad, probably because the assumed kinetic temperature is too high. We return to this issue in Sect. \ref{sect4profiles}.

\section{Discussion}\label{sect4}

We now discuss in more detail the results presented in the last Section, starting with the differences in homogeneous models arising from using different rate coefficients. We then discuss the abundance distributions in our hydrostatic core models, and conclude by looking into the simulated line emission spectra.

\begin{figure}
\resizebox{\hsize}{!}{\includegraphics{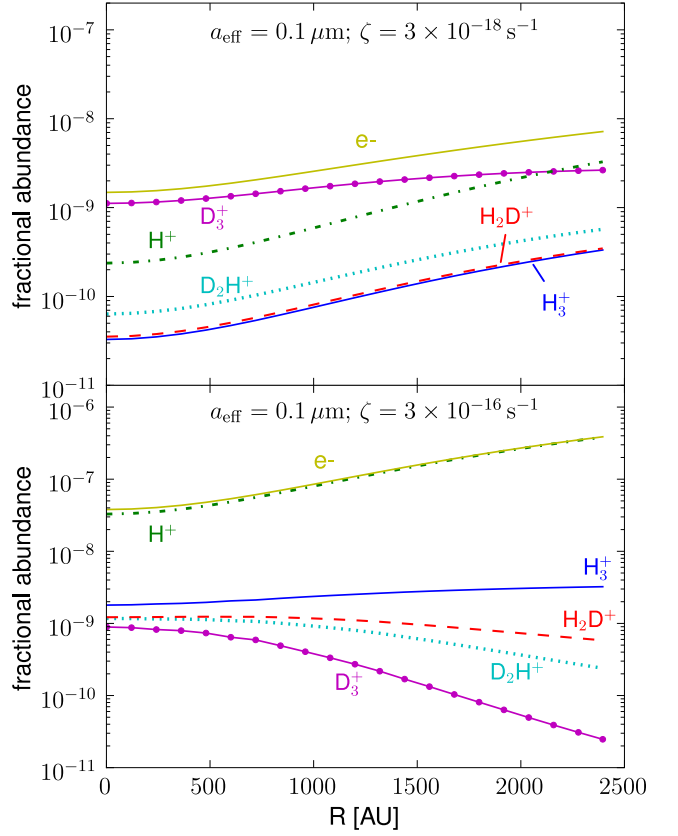}}
\caption{Same as the middle panel in Fig. \ref{fig: abundances_1.0}, but for cosmic ray ionization rates $\zeta = 3\times10^{-18}$\,s$^{-1}$ (upper panel) and $\zeta = 3\times10^{-16}$\,s$^{-1}$ (lower panel).}
\label{fig: crp_comparison}
\end{figure}

\subsection{Homogeneous models}

In the models calculated using the new reaction rate coefficients derived by \citet{HAS09}, the deuteration of ${\rm H_3^+}$ proceeds faster than in the models of \citet{FPW04}. As a consequence, D$_3^+$ becomes the dominant ion at relatively low densities. The total\footnote{By the "total rate coefficients" we mean the resultant forward and backward rate coefficients, $k_1$ and $k_{-1}$, in the presentation omitting the nuclear spin variants, e.g., ${\rm H_3^+} + {\rm HD} \mathop{\rightleftharpoons}\limits_{k_{-1}}^{k_1} {\rm H_2D^+} + {\rm H_2}$. In this presentation, the ${\rm H_2D^+}$ formation rate can be written $k_1 [{\rm H_3^+}][{\rm HD}]$, where $k_1$ is a combination of seven individual rate coefficients weighted according to the fractions of ${\rm H_3^+(o)}$ or ${\rm H_3^+(p)}$ of the total ${\rm H_3^+}$ abundance (see Table VIII of Hugo et al. 2009). The other deuteration reactions and their reverse reactions are composed of eight reactions. Besides the temperature, the total rate coefficients and the equilibrium constant, $K=k/k_{-}$, depend on the relative populations of nuclear spin variants, in particular o/p H$_2$ \citep[e.g.][]{Gerlich02}.} rate coefficients of \cite {HAS09} for the deuteration reactions ${\rm H_3^+} \rightarrow {\rm H_2D^+} \rightarrow {\rm D_2H^+} \rightarrow {\rm D_3^+}$ are larger than those used by \citet{FPW04} by an average factor of $\sim 4-5$. This difference originates in the discrepancy between the laboratory measurements of \citet{Gerlich02} and \citet[][see their Sect. IV.C]{HAS09} concerning the forward rate coefficients at very low temperatures. On the other hand, the total rate coefficients for the "backward" reactions ($k_{-1}$, $k_{-2}$, $k_{-3}$) are similar in the two reaction schemes. This can be traced to the fact that the "backward" rate coefficient for the reaction ${\rm H_2D^+}$(o) + ${\rm H_2}$(o) $\Arrow^{k_{-1}(oo)} {\rm H_3^+}$(o/p) + ${\rm HD}$ derived by \citet{HAS09} agrees with the estimate of \citet[][Sect. 3.3]{Gerlich02}.

The ratios $K=k/k_{-}$ are thus larger in \citet{HAS09} which implies that chemical equilibrium is established at higher concentrations of deuterated species.  The equilibrium constants in the present model are, however, smaller than in the models of \citet{Roberts04} and \citet{Caselli08}. The forward rate coefficients of \citet{HAS09} are similar to those used in the two previous studies while the backward rate coefficents are several orders of magnitude larger owing to the non-thermal H$_2$ (o/p) ratio.

As discussed in \cite{FPW06}, the {\sl ortho}/{\sl para} ratios of ${\rm H_3^+}$ and ${\rm H_2D^+}$ correlate with ${\rm H_2(o/p)}$ through proton exchange reactions. The H$_3^+$ ion is primarily produced in the {\sl para} form via the reaction of ${\rm H_2^+}$(p) with ${\rm H_2}$(p). The {\sl para-ortho} and {\sl ortho-para} conversions through reactions with ${\rm H_2}$(o) are, however, rapid, and compete only against the deuteration reaction with HD. In the conditions considered here a good approximation of the o/p ratio can be obtained from
$$
{\rm H_3^+(o/p)} \approx \frac{k_{po} \,{\rm H_2 (o/p)}}
                             {k_{1o} \, x({\rm HD}) + k_{op} \, {\rm H_2 (o/p)}} \; ,
$$ 
where $k_{po}$ and $k_{op}$ are the rate coefficients of the nuclear spin changing reactions of ${\rm H_3^+}$(p) and ${\rm H_3^+}$(o) with ${\rm H_2}$(o) (reactions 6 and 7 in Flower 2006), and $k_{1o}$ is the rate coefficient of the reaction of ${\rm H_3^+}$(o) with ${\rm HD}$ which produces mainly ${\rm H_2D^+}$(o) \citep[see Table VIII in][]{HAS09}. Using the \citet{HAS09} rate coefficients,  with $k_{po} \sim k_{op} \sim 0.3 k_{1o}$, we obtain a steady-state ratio  of ${\rm H_3^+(o/p)} \sim 0.2 - 0.4$ in cold, dense gas, where ${\rm H_2(o/p)} \sim x({\rm HD})$ (see Fig. \ref{fig: op_vs_t}).

Also for ${\rm H_2D^+}$, the nuclear spin changing reactions with ${\rm H_2}$ are effective and an approximate value of the o/p ratio can be obtained through balancing these according to Eq. (7) of \citet{Gerlich02}. Below $\sim 8$ K, however, primary production from ${\rm H_3^+} + {\rm HD}$ and destruction through the secondary deuteration reaction ${\rm H_2D^+} + {\rm HD} \rightarrow {\rm D_2H^+} + {\rm H_2}$ start to dominate.  At very low temperatures the ${\rm H_2D^+}$ (o/p) ratio settles somewhere between 2 and 3 (Fig. \ref{fig: op_vs_t}).

The production and destruction of {\sl ortho} and {\sl para} D$_2$H$^+$ occur mainly via the deuteration reactions ${\rm H_2D^+} + {\rm HD} \rightarrow {\rm D_2H^+} + {\rm H_2}$ and ${\rm D_2H^+} +
{\rm HD} \rightarrow {\rm D_3^+} + {\rm H_2}$.  As discussed in \cite{FPW06}, the lower energy {\sl ortho} state is favoured in the reverse reactions with ${\rm H_2(o)}$ \citep[reactions 11 and 12 in their
Sect. 3.4; Table VIII in][]{HAS09}.  Moreover, ${\rm D_2H^+(p)}$ is reduced through {\sl para-ortho} conversion in the reaction with HD. In the present model the p/o ratio is $\sim 0.2-0.4$ below 10 K and is not sensitive to $n({\rm H_2})$.

We note that while the $\rm H_2D^+$ (o/p) ratio is similar to that found in the models of \citet[][Fig. 9]{Pagani09}, the $\rm D_2H^+$ (p/o) ratio is larger by a factor of 2-4 in the present model. By performing some test runs we could establish that the main reason for this apparent discrepancy is in the different H$_2$ (o/p) ratio. In the time-dependent model of \citet{Pagani09}, the initial H$_2$ (o/p) ratio is larger than unity, and the chemical evolution is followed until the model matches with the constraints based on observations of the $\rm N_2D^+/N_2H^+$ ratio toward the L183 core.  The resulting H$_2$ (o/p) ratio ($\sim$0.005-0.05) is much higher than in our models (less than $10^{-4}$). At a high H$_2$ (o/p) ratio, $\rm D_2H^+$(p) is mainly destroyed in the reaction $\rm D_2H^+$(p) + H$_2$(o) $\rightarrow$ $\rm H_2D^+$(p) + HD, and the destruction by HD (including para-ortho conversion) is not significant.

The principal reactions determining the {\sl ortho/meta} ratio of ${\rm D_3^+}$ are discussed in \citet{FPW04} (Appendix B, with interchanged {\sl meta} and {\sl ortho} appellations with respect to the present work).  The (lowest energy) {\sl meta} form of ${\rm D_3^+}$ is primarily formed via deuteration of ${\rm D_2H^+(o)}$ by HD, and via {\sl ortho-meta} conversion in reaction ${\rm D_3^+(o)} + {\rm HD} \rightarrow {\rm D_3^+(m)} + {\rm HD}$.  The destruction of ${\rm D_3^+(m)}$ is overwhelmingly dominated by dissociative recombination with electrons or negatively charged grains. For ${\rm D_3^+(o)}$, the reaction ${\rm D_2H^+} + {\rm HD} \rightarrow {\rm D_3^+} + {\rm H_2}$ in the backward direction and the {\sl ortho-meta} conversion mentioned above are comparable with electron recombination as destruction mechanisms. The formula presented in the appendix of \citet{FPW04} approximates the equilibrium o/m ratio with a high accuracy.  The ratio is proportional to the electron density, and decreases as the gas density increases. On the other hand, it changes very little as a
function temperature in the range considered here (see Figs. 3 and 4).

Finally, we note that the ${\rm D_2}$ (p/o) ratio correlates with the ${\rm D_3^+}$ (o/m) ratio (see Fig.~3). The higher energy ${\rm D_2(p)}$ is principally formed from ${\rm D_3^+(o)}$ via dissociative recombination with free electrons and on dust grains, whereas ${\rm D_2(o)}$ forms primarily from ${\rm D_3^+(m)}$ (DR rate coefficients from Pagani et al. 2009 and branching ratios from Flower et al. 2004 have been used here). The destruction of ${\rm D_2(p)}$ is dominated by deuterium exchange reactions with ${\rm D_3^+(m)}$ and deuteration of ${\rm H_3^+(p)}$, while for ${\rm D_2(o)}$ the deuteration of ${\rm D_2H^+(o)}$ and ${\rm H_3^+(p)}$ are the most important destruction reactions. The resulting equilibrium ${\rm D_2}$ (p/o) ratio is roughly an order of magnitude lower than the ${\rm D_3^+}$ (o/m) ratio.

\subsection{Hydrostatic models}

In the hydrostatic models (Figs. \ref{fig: abundances_1.0} and \ref{fig: opratios}) combined effects of changing grain size, density and temperature can be seen. The decrease of the average grain size from the top panel to the bottom signifies an increase in the total grain surface area. This results in a diminishing fractional abundance of free electrons and a growing abundance of negatively charged grains (owing to the sticking of electrons onto neutral grains), and an increase of the ${\rm H_2}$ (o/p) ratio (because of enhanced ${\rm H_2}$ formation on grain surfaces).  The radial fall off of the density (from left to right) is accompanied by an elevated temperature at the outer edge of the core.

The ${\rm D_3^+}$ ion is the most abundant deuterated form of ${\rm H_3^+}$ in the centre of the core. In the following we use the ${\rm D_3^+}/{\rm H_3^+}$ abundance ratio to quantify the degree of deuteration. The ${\rm D_3^+}/{\rm H_3^+}$ ratio changes as a function of grain size. For the largest grain size, $a=0.2 \,\mu$m, $[{\rm D_3^+}]/[{\rm H_3^+}]$ is $\sim 3.3$ in the core center, and $\sim 0.08$ at the edge. As the grain size is decreased to $a=0.1 \,\mu$m,  $[{\rm D_3^+}]/[{\rm H_3^+}]$ increases both in the center (to $\sim 8.6$) and at the edge (to $\sim 0.6$). For still smaller grains, $a=0.05 \,\mu$m, $[{\rm D_3^+}]/[{\rm H_3^+}]$ drops slightly (to $8.5$) in the center, but increases to $\sim 1.4$ at the edge.

\begin{figure}
\resizebox{\hsize}{!}{\includegraphics{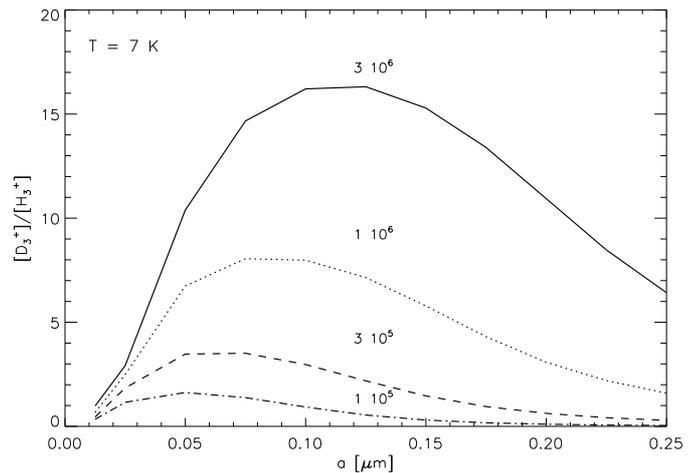}}
\caption{Dependence of the D fractionation on the average grain size. The curves correspond to different values of the gas density.}
\label{fig: dfrac}
\end{figure}

The changes in the deuterium fractionation as functions of the density and the effective grain radius are plotted in Fig. \ref{fig: dfrac} for isothermal models with $T_{\rm kin} = 7$ K. One can see in the plot that for each value of density, there is a deuteration maximum which shifts to lower values of $a$ as the density is decreased. The sloping down of $[{\rm D_3^+}]/[{\rm H_3^+}]$ on the right, towards larger grains, is caused by an increasing abundance of free electrons.  The decrease of $[{\rm D_3^+}]/[{\rm H_3^+}]$ on the left side, towards smaller grains, is caused by the increase in the abundance of H$_2$(o) which hinders deuteration by converting ${\rm H_2D^+}$ back to ${\rm H_3^+}$, and by the increase of negatively charged grains which replace electrons as the principal destructors of ${\rm D_3^+}$.

Figure \ref{fig: crp_comparison} shows radial abundance profiles in a chemical model otherwise similar to the middle panel of Fig. \ref{fig: abundances_1.0}, but for cosmic ionization rates $\zeta=3\times10^{-18}$ s$^{-1}$ and $\zeta=3\times10^{-16}$ s$^{-1}$.  It can be immediately seen in Fig. \ref{fig: crp_comparison}, as compared with the middle panel in Fig. \ref{fig: abundances_1.0}, that the abundances of ${\rm H^+}$ and ${\rm H_3^+}$ follow the change in the cosmic ray ionization rate. The H$^+$ abundance is proportional to $\zeta$ via dissociation of H$_2$ by cosmic rays and via dissociative charge transfer reaction between He$^+$ and H$_2$ \citep[see, e.g.][Appendix A]{WFP04}. The formation of ${\rm H_3^+}$ depends on $\zeta$ through ${\rm H_2^+}$, which is the principal product of the collision between ${\rm H_2}$ and a cosmic ray proton \citep[e.g.,][]{WFP04, Pagani09}. The ${\rm H_2D^+}$ and ${\rm D_2H^+}$ abundances are dragged upwards with ${\rm H_3^+}$ with an intensified cosmic ray ionization, whereas the ${\rm D_3^+}$ abundance remains roughly constant in the core center and falls off in the outer regions. So, quite reasonably, the $[{\rm D_3^+}]/[{\rm H_3^+}]$ ratio which depends on DR reactions diminishes strongly with an increasing $\zeta$, whereas the effects on $[{\rm H_2D^+}]/[{\rm H_3^+}]$ and $[{\rm D_2H^+}]/[{\rm H_3^+}]$ which depend on HD and ${\rm H_2(o/p)}$, are less notable.

\begin{figure}
\resizebox{\hsize}{!}{\includegraphics{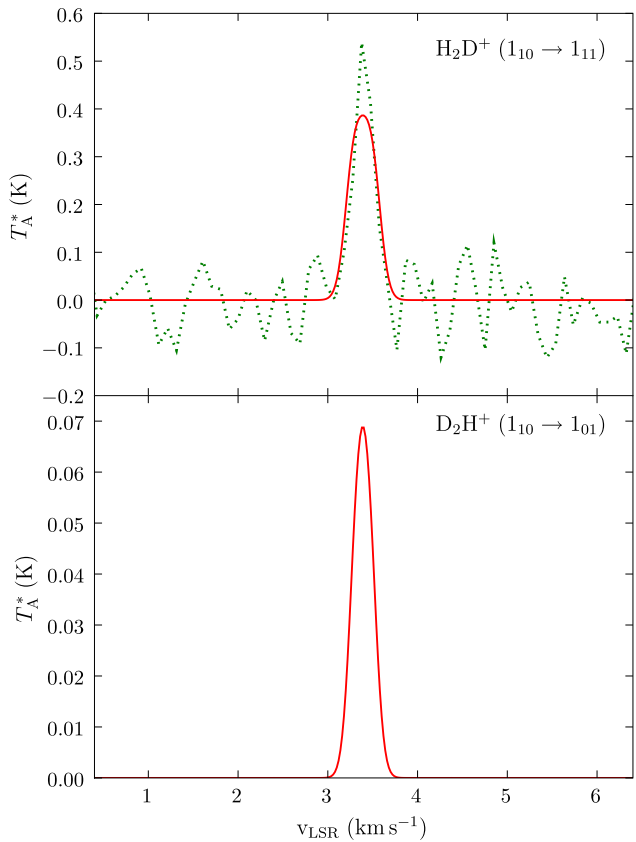}}
\caption{Predicted line profiles for the H$_2$D$^+$ $1_{10}$ -- $1_{11}$ (ortho) and the D$_2$H$^+$ $1_{10}$ -- $1_{01}$ (para) transitions as observed with APEX for the best-fit core model (solid curves). Also plotted is the observation toward Oph~D by \citet{Harju08} for the H$_2$D$^+$ $1_{10}$ -- $1_{11}$ transition (dotted curve).}
\label{fig: spectra}
\end{figure}

\begin{figure}
\resizebox{\hsize}{!}{\includegraphics{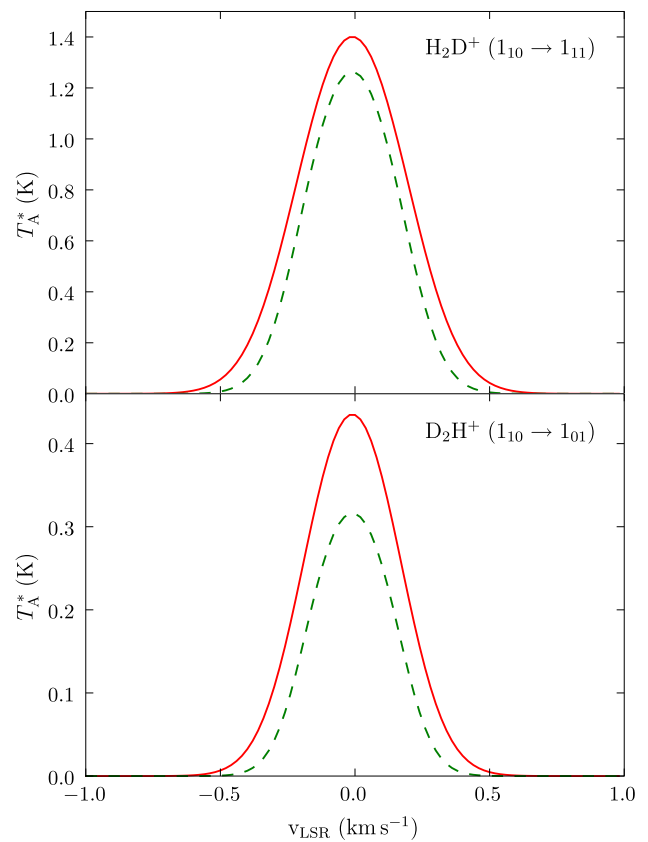}}
\caption{Predicted molecular line profiles for the H$_2$D$^+$ ($1_{10}$ -- $1_{11}$) and the D$_2$H$^+$ ($1_{10}$ -- $1_{01}$) transitions for the prestellar core 16293E as observed with the CSO, using the core parameters of \citet{Stark04} (solid curve). The modified model (see text) is also plotted (dashed curve).}
\label{fig: 16293_spectra}
\end{figure}

\subsection{Molecular line profiles}\label{sect4profiles}

The observed ${\rm H_2D^+}$(o) profile from Oph D can be reproduced reasonably well using a hydrostatic core model together with the "complete depletion" chemistry model with the rate coefficients of \citet{HAS09} (see Fig. \ref{fig: spectra}). The model giving the best agreement with the observed ${\rm H_2D^+}$(o) spectrum (see Sect. \ref{sect3spectra}) predicts a very weak ${\rm D_2H^+}$(p) line at 692 GHz ($T_{\rm A}^* \sim 0.07$ K with APEX). This line has a lower peak opacity ($\tau \sim 0.6$) than the ${\rm H_2D^+}$(o) line, whereas the radial $T_{\rm ex}$ distributions of the two lines are similar. One should note that even an optically thick ${\rm D_2H^+}$(p) $(1_{10}-1_{01})$ line from a cold source is weak in the brightness temperature scale because the frequency lies far away from the range where the Rayleigh-Jeans approximation is valid.

The physical model adopted from Stark et al. (2004) (see Sect. \ref{sect3spectra}) for the core I16293E produces easily the bright ${\rm H_2D^+}$(o) and ${\rm D_2H^+}$(p) lines observed by \citet{Vastel04} (see Fig. \ref{fig: 16293_spectra}, solid curves). The relatively high temperature (16 - 20 K) assumed by \citet{Stark04} gives, however, rise to too broad lines as compared with the observations. Another problem with this assumption is that the chemistry model predicts a clearly higher D$_2$H$^+$(p)/H$_2$D$^+$(o) abundance ratio around 16 K \citep[see Fig. 7 in][]{FPW04}. We carried out another simulation assuming that the kinetic temperature is 10 K in the center and that it increases gradually to 12 K at the edge. The fractional H$_2$D$^+$(o) and D$_2$H$^+$(p) abundances were assumed to be roughly equal, in agreement with the prediction of the chemistry model. The modelled spectra are shown in Fig. \ref{fig: 16293_spectra} (dashed curves). Now both intensities and line widths are within the range reported by \citet{Vastel04}. The H$_2$D$^+$(o) and D$_2$H$^+$(p) column densities, both $\sim$ $2.3\times10^{-13} \, {\rm cm}^{-2}$, are slightly higher than those derived by \citet{Vastel04}.

\section{Conclusions}\label{sect5}

We studied the chemistry in molecular cloud cores in the complete depletion limit, taking advantage of new state-to-state chemical reaction rate coefficients for the H$_3^+$ + H$_2$ reacting system \citep{HAS09}. We used the modelling results to predict line emission spectra for the H$_2$D$^+$ $1_{10}$ -- $1_{11}$ (372 GHz) and the D$_2$H$^+$ $1_{10}$ -- $1_{01}$ (692\,GHz) transitions in a core model corresponding to the prestellar core Oph D. We also carried out simulations of the above transitions for the I16293E core. The simulated spectra were compared with observations by \citet{Harju08} and \citet{Vastel04} for the respective cores.

In this paper, we used steady-state chemical abundances in the molecular line simulations. It has been noted in the literature that the timescale for chemical development can be an order of magnitude larger than the dynamical timescale, and thus chemical steady-state might not be reached before collapse begins \citep[see for example][]{FPW06}. In general, however, ions reach chemical steady-state earlier than neutral species, due to the speed of the ion-molecule reactions. In fact, our chemical models indicate that ions reach steady-state before $t = 1\times10^5$ years. On the other hand, the relatively high $\rm H_2D^+$(o) and $\rm D_2H^+$(p) column densities derived toward Oph D and I16293E imply an appreciable degree of deuteration, which suggests that the cores are in an advanced chemical state. Taking into account these arguments, the assumption of steady-state seems, in the present context, appropriate.

The simulated profile of the H$_2$D$^+$ ($1_{10}$ -- $1_{11}$) line from a self-consistent hydrostatic core model (including chemistry modelling) is in reasonably good agreement with observations toward
Oph D. The model predicts that the D$_2$H$^+$ ($1_{10}$ -- $1_{01}$) line at 692 GHz probably cannot be detected in this source. In fact, this transition is difficult to detect in emission in any cold object
because of its high frequency. The same is true for the H$_2$D$^+$(p) and D$_2$H$^+$(o) ground state lines, which lie at even higher frequencies of $\sim$ 1.4\,THz.

The Oph D core seems particularly appropriate for using a hydrostatic model. However, the chemistry model used here begins to lose its validity in the outer parts, where the density drops below $10^6$\,cm$^{-3}$. In these conditions one can no longer ignore heavier substances in the gas phase, which modify the abundances of ${\rm H_3^+}$ and its deuterated forms. So, in reality, the ${\rm H_2D^+}$ and ${\rm D_2H^+}$ abundances which in our models rise toward the core edge are likely to turn down outside the dense nucleus. This effect can be seen in the models of \citet{Pagani09}. It is obvious that chemical modelling should be extended to account for the presence of heavier species. This would also justify studies of larger cores, presenting larger density gradients.

Our simulations of molecular line profiles toward the I16293E core are in good agreement with the observational results presented in \citet{Vastel04}. While line emission simulations utilizing the physical core parameters of \citet{Stark04} produce enough emission, we found that to produce linewidths comparable to those in \citet{Vastel04}, the core temperature profile and chemical abundances should be modified. This modification is also justified by results from chemical modelling. The H$_2$D$^+$(o) and D$_2$H$^+$(p) column densities are found to become comparable when the temperature drops to $\sim$ 10 K.

The deuteration of H$_3^+$ proceeds much faster in the present reaction scheme using the \citet{HAS09} results than in the \citet{FPW04} model where the deuteration rate coefficients were adopted from \citet{Gerlich02}. As a consequence, the steady-state $\rm D_3^+/H_3^+$ abundance ratio becomes larger than in the models of Flower, Pineau des Forêts \& Walmsley. However, it should be noted that already in \citet{WFP04}, D$_3^+$ becomes the most abundant isotopologue of H$_3^+$ at very high densities, and that this phenomenon was predicted previously by \citet{Roberts03}.

The low-temperature rate coefficients from \citet{Gerlich02} are lower than those by \citet{HAS09} by a factor of $\sim$ 4. The experimental circumstances have been thoroughly discussed in \citet{HAS09}, concluding that further experimental investigations, preferably with a different setup, are urgently needed. The ortho/para ratios of the deuterated forms of H$_3^+$ are also modified in the new model. Along with the changes in abundances, this has an impact on the observability of H$_2$D$^+$ and D$_2$H$^+$ toward prestellar cores. We find at very low temperatures a lower H$_2$D$^+$ (o/p) ratio, and a slightly higher D$_2$H$^+$ (p/o) ratio than predicted by using the coefficients adopted in \citet{FPW04}.

The state-to-state coefficients calculated by \citet{HAS09} provide an opportunity to refine both the chemistry model and the model for the excitation of the rotational transitions of interest. Besides collisions with {\sl para} and {\sl ortho} H$_2$, it would seem reasonable to examine the excitation of rotational levels of ${\rm H_2D^+}$ and ${\rm D_2H^+}$ in collisions with HD. The next step forward is to combine the full state-to-state reaction network with radiative transfer calculations. The realization of this
improvement looks particularly feasible for the complete depletion model.

\begin{acknowledgements}
O.S. and J.H. acknowledge support from the Academy of Finland grant no. 117206. We would also like to thank D. Flower, G. Pineau des Forêts, C.M. Walmsley and L. Pagani for their comments on this paper, and the anonymous referee for his/her helpful report.
\end{acknowledgements}
\bibliographystyle{aa}
\bibliography{13350ref}
\end{document}